\documentclass[conference,10pt]{IEEEtran}
\IEEEoverridecommandlockouts
\usepackage{amsmath,amssymb,amsfonts}
\usepackage{orcidlink}
\usepackage{mathtools}
\usepackage{bm}
\usepackage{algorithmic}
\usepackage{hyperref}
\hypersetup{hidelinks}
\usepackage{graphicx}
\usepackage{textcomp}
\usepackage{xcolor}
\usepackage[acronym,toc]{glossaries}
\glsdisablehyper
\usepackage{cleveref}
\usepackage{siunitx}
\usepackage{tabularx}
\usepackage{booktabs}
\usepackage{pgfplots, tikz}
\pgfplotsset{compat=1.17}
\usetikzlibrary{arrows.meta}
\usepackage{ifthen}
\usepackage{environ}
\usepackage{placeins}
\usepackage{csquotes}
\ifCLASSOPTIONcompsoc
    \usepackage[caption=false, font=normalsize, labelfont=sf, textfont=sf]{subfig}
\else
    \usepackage[caption=false, font=footnotesize]{subfig}
\fi
\usepackage[activate={true,nocompatibility},
    final,
    tracking=true,
    kerning=true,
    spacing=true,
    factor=1100,
    stretch=25,
    shrink=25,
    letterspace=-30
]{microtype}
\newcommand{\ninept}{\fontsize{9pt}{10.5pt}\selectfont}
\ninept
\makeatletter
\newsavebox{\measure@tikzpicture}
\NewEnviron{scaletikzpicturetowidth}[1]{%
    \def\tikz@width{#1}%
    \begin{lrbox}{\measure@tikzpicture}%
        \BODY
    \end{lrbox}%
    \pgfmathparse{#1/\wd\measure@tikzpicture}%
    \BODY
}
\makeatother

\newacronym{admm}{ADMM}{Alternating Directions of Multipliers Method}
\newacronym{anm}{ANM}{Atomic Norm Minimization}
\newacronym{adc}{ADC}{Analog-to-Digital Converter}
\newacronym{awgn}{AWGN}{Additive White Gaussian Noise}
\newacronym{asic}{ASIC}{Application Specific Integrated Circuit}
\newacronym{arpack}{ARPACK}{ARnoldi PACKage}
\newacronym{api}{API}{Application Programmable Interface}
\newacronym{aut}{AUT}{Antenna Under Test}
\newacronym[plural=AOI, firstplural=Areas of Interest (AOI)]{aoi}{AOI}{Area of Interest}
\newacronym{ai}{AI}{Artifical Intelligence}
\newacronym{aic}{AIC}{Akaike Information Criterion}

\newacronym{bp}{BP}{Basis Pursuit}
\newacronym{bpdn}{BPDN}{Basis Pursuit Denoising}
\newacronym{blue}{BLUE}{Best Linear Unbiased Estimator}

\newacronym{cnn}{CNN}{Convolutional Neural Network}
\newacronym{crb}{CRB}{Cramér-Rao Bound}
\newacronym{cs}{CS}{Compressed Sensing}
\newacronym{cf}{CF}{Crest factor}
\newacronym{cr}{CR}{Compression Ratio}
\newacronym{cmos}{CMOS}{Complementary Metal Oxide Semiconductor}
\newacronym{cots}{COTS}{Commercial-Off-The-Shelf}
\newacronym{cpu}{CPU}{Central Processing Unit}
\newacronym{cfar}{CFAR}{Constant False Alarm Rate}
\newacronym{cvx}{CVX}{Convex Optimization toolboX}
\newacronym{cp}{CP}{Cyclic Prefix}

\newacronym{doa}{DoA}{Direction of Arrival}
\newacronym{dod}{DoD}{Direction of Departure}
\newacronym[plural=DMC,firstplural=Dense Multipath Components (DMC)]{dmc}{DMC}{Dense Multipath Components}
\newacronym{das}{DAS}{Delay-and-Sum}
\newacronym{dft}{DFT}{Discrete Fourier Transform}
\newacronym{dvbt}{DVB-T}{}
\newacronym{idft}{IDFT}{Inverse Discrete Fourier Transform}
\newacronym{dct}{DCT}{Discrete Cosine Transform}
\newacronym{dsp}{DSP}{Digital Signal Processor}
\newacronym{dnn}{DNN}{Deep Neural Network}
\newacronym{dpss}{DPSS}{Discrete Prolate Spheroidal Sequences}

\newacronym{eadf}{EADF}{Effective Aperture Distribution Function}
\newacronym{etadf}{ETADF}{Effective Time-Aperture Distribution Function}
\newacronym{esprit}{ESPRIT}{Estimation of Signal Parameters via Rotational Invariance Techniques}
\newacronym{ett}{ETT}{Eigenvalue Threshold Test}
\newacronym{edc}{EDC}{Efficient Detection Criterion}
\newacronym{eft}{EFT}{Exponential Fitting Test}

\newacronym{fc}{FC}{fully-connected}
\newacronym{fht}{FHT}{Fast Hadamard Transform}
\newacronym[longplural={Fast Fourier Transforms}]{fft}{FFT}{Fast Fourier Transform}
\newacronym{fmcw}{FMCW}{Frequency-Modulated Continuous-Wave}
\newacronym{fpga}{FPGA}{Field Programmable Gate Array}
\newacronym{fri}{FRI}{Finite Rate of Innovation}
\newacronym{fim}{FIM}{Fisher Information Matrix}
\newacronym{fmc}{FMC}{Full Matrix Capture}
\newacronym{fista}{FISTA}{Fast Iterative Shrinkage-Thresholding Algorithm}
\newacronym{frvm}{FRVM}{Fast Relevance Vector Machine}

\newacronym{gfcs}{grid-free CS}{grid-free compressive sensing}
\newacronym{gpu}{GPU}{Graphical Processing Unit}
\newacronym{gtd}{GTD}{Geometrical Theory of Diffraction}
\newacronym{gan}{GAN}{Generative Adversarial Network}

\newacronym{hrpe}{HRPE}{High Resolution Parameter Estimator}
\newacronym{hfss}{Ansys HFSS}{Ansys High Frequency Electromagnetic Simulation Software}

\newacronym[longplural={Inverse Fast Fourier Transforms}]{ifft}{IFFT}{Inverse Fast Fourier Transform}
\newacronym{ir}{IR}{Impulse Response}
\newacronym{iid}{iid}{independent and identically distributed}
\newacronym{iir}{IIR}{Infinite Impulse Response}
\newacronym{irf}{IRF}{Impulse Response Function}
\newacronym{ista}{ISTA}{Iterative Shrinkage-Thresholding Algorithm}
\newacronym{isac}{ISAC}{Integrated Sensing and Communication}

\newacronym{kld}{KLD}{Kullback-Leibler Divergence}

\newacronym{ls}{LS}{Least Squares}
\newacronym{lasso}{LASSO}{Least Absolute Shrinkage and Selection Operator}
\newacronym{lse}{LSE}{Line Spectral Estimation}
\newacronym{lfsr}{LFSR}{Linear Feedback Shift Register}
\newacronym{lo}{LO}{Local Oscillator}
\newacronym{los}{LOS}{Line of Sight}
\newacronym{lti}{LTI}{Linear Time-Invariant}
\newacronym{ltv}{LTV}{Linear Time-Variant}
\newacronym{lam}{LAM}{Large Area Monitoring}

\newacronym{mimo}{MIMO}{multiple input multiple output}
\newacronym{mmv}{MMV}{multiple measurement vectors}
\newacronym{ml}{ML}{Machine Learning}
\newacronym{mmse}{MMSE}{misspecified mean squared error}
\newacronym{mse}{MSE}{mean squared error}
\newacronym{bce}{BCE}{Binary Crossentropy}
\newacronym{mlbs}{MLBS}{Maximum Length Binary Sequence}
\newacronym{mpc}{MPC}{Multipath Component}
\newacronym{msm}{MSM}{M-Sequence Method}
\newacronym{mwc}{MWC}{Modulated Wideband Converter}
\newacronym{mpm}{MPM}{Matrix Pencil Method}
\newacronym{mpu}{MPU}{Microprocessor Unit}
\newacronym{mri}{MRI}{Magnetic resonance imaging}
\newacronym{music}{MUSIC}{Multiple Signal Classification}
\newacronym{mkl}{MKL}{Math Kernel Library}
\newacronym{mcrb}{MCRB}{Misspecified Cramér-Rao Bound}
\newacronym{mmle}{MMLE}{Misspecified Maximum-Likelihood Estimator}
\newacronym{mtt}{MTT}{Multi-Target Tracking}

\newacronym{ndt}{NDT}{Nondestructive Testing}
\newacronym{nde}{NDE}{Nondestructive Evaluation}
\newacronym{nn}{NN}{Neural Net}
\newacronym{nist}{NIST}{National Institute of Standards and Technology}

\newacronym{omp}{OMP}{Orthogonal Matching Pursuit}
\newacronym{oop}{OOP}{Object Oriented Programming}
\newacronym{ofdm}{OFDM}{Orthogonal Frequency-Division Multiplexing}

\newacronym{pdp}{PDP}{Power Delay Profile}
\newacronym{pn}{PN}{Pseudo-Noise}
\newacronym{pwc}{PWC}{Plane Wave Compounding}
\newacronym{pwi}{PWI}{Plane Wave Imaging}
\newacronym{pura}{PURA}{Patch Uniform Rectangular Array}
\newacronym{pymax}{PyMAX}{Python Maximization Approach}
\newacronym{pts}{PTS}{Pseudo-True Solution}
\newacronym{pdf}{pdf}{probability density function}
\newacronym{plmn}{PLMN}{Public Land Mobile Network}
\newacronym{ppf}{PPF}{Pulse-Pair Filter}
\newacronym{piml}{PIML}{Physics-Informed Machine Learning}

\newacronym{relu}{ReLU}{Rectified Linear Unit}
\newacronym{resnet}{ResNet}{Residual Neural Network}
\newacronym{ram}{RAM}{Random Access Memory}
\newacronym{rcs}{RCS}{Radar Cross Section}
\newacronym{rd}{RD}{Random Demodulator}
\newacronym{rx}{Rx}{receiver}
\newacronym{rem}{REM}{Reconstruction Error Metric}
\newacronym{rmse}{RMSE}{Root Mean Squared Error}
\newacronym{rms}{RMS}{root mean squared}
\newacronym{ric}{RIC}{Restricted Isometry Constant}
\newacronym{rip}{RIP}{Restricted Isometry Property}
\newacronym{rc}{RC}{Raised Cosine}
\newacronym{roi}{ROI}{Region of Interest}
\newacronym{rimax}{RIMAX}{Richter Maximization Approach}
\glsunset{rimax}
\newacronym{rvm}{RVM}{Relevance Vector Machine}
\newacronym{rru}{RRU}{Remote Radio Unit}
\newacronym{rtk}{RTK}{Real-time kinematic}

\newacronym{scf}{SCF}{spatial correlation function}
\newacronym{samurai}{SAMURAI}{Synthetic Aperture Measurements of Uncertainty in Angle of Incidence}
\newacronym[plural=SC,firstplural=Specular Components (SC)]{sc}{SC}{Specular Components}
\newacronym{sdp}{SDP}{semi-definite program}
\newacronym{sdr}{SDR}{Signal to Diffuse Ratio}
\newacronym{simd}{SIMD}{Single Instruction Multiple Data}
\newacronym{svd}{SVD}{singular value decomposition}
\newacronym{svm}{SVM}{Support Vector Machine}
\newacronym{soe}{SOE}{Sparsity Order Estimation}
\newacronym{sgd}{SGD}{Stochastic Gradient Descent}
\newacronym{stuca}{StUCA}{Stacked Uniform Circular Array}
\newacronym{spucpa}{SPUCPA}{Stacked Polarimetric Uniform Circular Patch Array}
\newacronym{suca}{SUCA}{Stacked Uniform Circular Array}
\newacronym{saft}{SAFT}{Synthetic Aperture Focusing Technique}
\newacronym{sota}{SOTA}{State of the Art}
\newacronym{ssd}{SSD}{Solid State Device}
\newacronym{ssr}{SSR}{Sparse Signal Recovery}
\newacronym{sa}{SA}{Synthetic Aperture}
\newacronym{spw}{SPW}{Single Plane Wave}
\newacronym{shm}{SHM}{Structural Health Monitoring}
\newacronym{snr}{SNR}{Signal-to-Noise Ratio}
\newacronym{stela}{STELA}{Soft-Thresholding with Exact Line Search Algorithm}
\newacronym{siso}{SISO}{Single Input Single Output}
\newacronym{simo}{SIMO}{Single Input Multiple Output}
\newacronym{swe}{SWE}{Spherical Wave Expansion}
\newacronym{sme}{SME}{Spherical Mode Expansion}
\newacronym{sage}{SAGE}{Space-Alternating Generalized Expectation-Maximization}

\newacronym{tdd}{TDD}{Time-division duplexing}
\newacronym{th}{T\&H}{Track and Hold}
\newacronym{tf}{TF}{Transfer Function}
\newacronym{tx}{Tx}{transmitter}
\newacronym{twista}{TWISTA}{Two-step Iterative Shrinkage-Thresholding Algorithm}
\newacronym{tof}{ToF}{Time-of-Flight}
\newacronym{tdoa}{TDOA}{Time-Difference of Arrival}

\newacronym{uca}{UCA}{uniform circular array}
\newacronym{ura}{URA}{Uniform Rectangular Array}
\newacronym{ula}{ULA}{Uniform Linear Array}
\newacronym{uwb}{UWB}{Ultra-Wideband}
\newacronym{usndt}{US-NDT}{Ultrasonic Non-destructive Testing}
\newacronym{uav}{UAV}{Unmanned Aerial Vehicle}
\newacronym{utm}{UTM}{Unmanned Aircraft Traffic Management}

\newacronym{vna}{VNA}{Vector Network Analyser}
\newacronym{vsh}{VSH}{Vector Spherical Harmonics}
\def\BibTeX{{\rm B\kern-.05em{\sc i\kern-.025em b}\kern-.08em
T\kern-.1667em\lower.7ex\hbox{E}\kern-.125emX}}

\newcommand{\N}{\mathbb{N}}
\newcommand{\R}{\mathbb{R}}
\newcommand{\C}{\mathbb{C}}

\definecolor{QC1}{HTML}{0077BB}
\definecolor{QC2}{HTML}{33BBEE}
\definecolor{QC3}{HTML}{009988}
\definecolor{QC4}{HTML}{EE7733}
\definecolor{QC5}{HTML}{CC3311}
\definecolor{QC6}{HTML}{EE3377}
\definecolor{QC7}{HTML}{BBBBBB}

\newcommand{\reddot}{\tikz{\fill[QC5, draw=black] (0,0) circle (2pt);}}
\newcommand{\bluedot}{\tikz{\fill[QC1, draw=black] (0,0) circle (2pt);}}
\newcommand{\dashedline}{\tikz[baseline=-0.5ex]{\draw[dashed] (0,0) -- (1em,0);}}

\newcommand{\solidline}{\tikz[baseline=-0.5ex]{\draw[solid] (0,0) -- (1em,0);}}

\ifx\qty\undefined
    \let\oldnum\num
    \renewcommand{\num}[2][ignored]{\oldnum{#2}}

    \ifx\qty\undefined
        \newcommand{\qty}[3][ignored]{\SI{#2}{#3}}
    \fi

    \ifx\qtyrange\undefined
        \newcommand{\qtyrange}[4][ignored]{\SIrange{#2}{#3}{#4}}
    \fi
\fi

\usepackage[
    sorting=none,
    maxbibnames=99,
    defernumbers=true,
    style=ieee,
    giveninits=true,
    maxnames=3,
    url=false,
]{biblatex}
\AtEveryBibitem{
    \clearlist{address}
    \clearfield{url}
    \clearfield{month}
    \clearfield{editor}
    \clearfield{eprint}
    \clearfield{volume}
    \clearfield{isbn}
    \clearfield{issn}
    \clearfield{pages}
    \clearfield{addendum}
    \clearlist{location}
    \clearlist{pages}
    \clearlist{editor}
}
\makeatletter
\let\blx@rerun@biber\relax
\makeatother

\usepackage{pgfplots}
\pgfplotsset{compat=1.17}
\usepackage{tikz}
\usetikzlibrary{arrows}
\usetikzlibrary{arrows.meta}
\usetikzlibrary{shapes}
\usetikzlibrary{shapes.geometric}
\usetikzlibrary{shapes.misc}
\usetikzlibrary{decorations.text}

\pgfdeclarelayer{edgelayer}
\pgfdeclarelayer{nodelayer}
\pgfsetlayers{background,edgelayer,nodelayer,main}

\usepackage{tikz}
\usetikzlibrary{backgrounds}
\usetikzlibrary{arrows}
\usetikzlibrary{shapes,shapes.geometric,shapes.misc}

\tikzstyle{tikzfig}=[baseline=-0.25em,scale=0.5]

\pgfkeys{/tikz/tikzit fill/.initial=0}
\pgfkeys{/tikz/tikzit draw/.initial=0}
\pgfkeys{/tikz/tikzit shape/.initial=0}
\pgfkeys{/tikz/tikzit category/.initial=0}

\pgfdeclarelayer{edgelayer}
\pgfdeclarelayer{nodelayer}
\pgfsetlayers{background,edgelayer,nodelayer,main}

\tikzstyle{none}=[inner sep=0mm]

\newcommand{\tikzfig}[1]{%
{\tikzstyle{every picture}=[tikzfig]
\IfFileExists{#1.tikz}
  {\input{#1.tikz}}
  {%
    \IfFileExists{./figures/#1.tikz}
      {\input{./figures/#1.tikz}}
      {\tikz[baseline=-0.5em]{\node[draw=red,font=\color{red},fill=red!10!white] {\textit{#1}};}}%
  }}%
}

\tikzstyle{every loop}=[]

\glsenableentrycount
\let\gls\cgls

\Crefname{figure}{Fig.}{Figs.}
\Crefname{tabular}{Tab.}{Tabs.}

\def\BibTeX{{\rm B\kern-.05em{\sc i\kern-.025em b}\kern-.08em
    T\kern-.1667em\lower.7ex\hbox{E}\kern-.125emX}}

\begin{document}

% ISAC, UAV, CNN-based
\title{Measurement-based Evaluation of CNN-based Detection and Estimation for ISAC Systems}

\author{\IEEEauthorblockN{
Steffen Schieler\IEEEauthorrefmark{1}\orcidlink{0000-0003-4480-234X}, 
Sebastian Semper\IEEEauthorrefmark{1}\orcidlink{0000-0002-2610-7389},
Christian Schneider\IEEEauthorrefmark{1}\orcidlink{0000-0003-1833-4562},
Reiner Thom\"a\IEEEauthorrefmark{1}\orcidlink{0000-0002-9254-814X}
}                                     % ...
%\\
\IEEEauthorblockA{\IEEEauthorrefmark{1}% 1st affiliations
Technische Universit\"at Ilmenau: FG EMS, Ilmenau, Germany, steffen.schieler@tu-ilmenau.de}
\IEEEauthorblockA{\IEEEauthorrefmark{2}% 2nd affiliations
Fraunhofer Institute of Integrated Circuits: Dep. EMS, Ilmenau, Germany}  
\thanks{\textcopyright 2025 IEEE. Personal use of this material is permitted. Permission from IEEE must be obtained for all other uses, in any current or future media, including reprinting/republishing this material for advertising or promotional purposes, creating new collective works, for resale or redistribution to servers or lists, or reuse of any copyrighted component of this work in other works.}
\thanks{This is the author’s accepted manuscript. The published version is available in IEEE Xplore under DOI 10.1109/RADAR52380.2025.11031730.}
}

\maketitle

\begin{abstract}
In wireless sensing applications, such as \gls{isac}, one of the first crucial signal processing steps is the detection and estimation targets from a channel estimate. 
Effective algorithms in this context must be robust across a broad \gls{snr} range, capable of handling an unknown number of targets, and computationally efficient for real-time implementation. 
During the last decade, different Machine Learning methods have emerged as promising solutions, either as standalone models or as complementing existing techniques. 
However, since models are often trained and evaluated on synthetic data from existing models, applying them to measurement is challenging. 
All the while, training directly on measurement data is prohibitive in complex propagation scenarios as a groundtruth is not available. 
Therefore, in this paper, we train a \gls{cnn} approach for target detection and estimation on synthetic data and evaluate it on measurement data from a suburban outdoor measurement. 
Using knowledge of the environment as well as available groundtruth positions, we study the detection probability and accuracy of our approach. 
The results demonstrate that our approach works on measurement data and is suitable for joint detection and estimation of sensing targets in \gls{isac} systems.
\end{abstract}


\section{Introduction}\label{section:1}
In recent years, \gls{ml} techniques have been investigated for different signal processing tasks in radar sensing.
One such task is the detection and estimation of target parameters from a wireless channel impulse response, typically including \gls{doa}, delay, and Doppler-shifts, e.g., in ~\cite{chen_robust_2022,liu_direction--arrival_2018,papageorgiou_deep_2021,shmuel_subspacenet_2024,schieler_grid-free_2024}.
Specifically for \gls{isac}, target detection and estimation present new challenges, in large part because the sensing is performed with systems and signals originally optimized for communication, which are typically based on \gls{ofdm}.
As \gls{isac} targets multi-user systems, it presents unique challenges compared to broadcast-based \gls{ofdm} - radar sensing systems, such as \gls{dvbt}.
For example, to accommodate multiple users simultaneously, these systems employ various scheduling procedures, which ultimately result in signals with varying degrees of sparsity in time, frequency, and space, which affect the shape of the ambiguity function~\cite{thoma_cooperative_2019}.
Similarly~\cite{tosi_feasibility_2024} noted, switching between down- and uplink in \gls{tdd} systems introduces ghost targets, increasing false alarm rates.
Furthermore, the different modulation orders can amplify noise in the channel estimate of subcarriers with small or no power allocated, resulting in increased false-alarm rates, e.g., requiring \gls{isac}-specific constellation shaping~\cite{geiger_joint_2025}.  
To meet these challenges, novel signal processing approaches are studied and \gls{ml}-based solutions are often singled out for their good performance and flexibility~\cite{shatov_joint_2024}. 

Notably, most of the approaches based on \gls{ml} explored rely on supervised learning, and hence, the availability of labeled data.
A straightforward approach is to obtain labeled data through measurements using an existing estimation algorithm. 
However, this method is problematic \gls{isac} due to several factors. 
First, measurements are costly and introduce a dependency on the statistics of the propagation environment, e.g., the distribution of target parameters in an urban crossroad scenario is different from a suburban industrial area.
This means, if only measurements from a single site are used, the \gls{ml} solution is likely to overfit and perform worse when applied in another scenario, which would mean that site-specific measurements are required, driving up cost.
Secondly, measurements capture not just the wireless channel propagation, but also the (possibly nonlinear) distortion effects of the measurement hardware. Removing them requires calibration.
While this is feasible for single measurements, the resulting mismatch is still encountered \gls{isac}-deployments, where equipment from different vendors must interoperate. 

In contrast, supervised models can be developed with \enquote{synthetic data}, generated from a suitable signal model, in a form of \gls{piml}. 
By selecting and tuning the parameters of the model, the approach allows full control over the properties of the generated data, from the distribution of targets up to the level of distortions introduced from the measurement equipment.
However, it is necessary to evaluate the obtained models on real measurement data, as evaluation with synthetic data can unintentionally mask inappropriate assumptions made in the signal model and lead to wrong performance assessments.
For example, a \gls{ml} model trained to predict a fixed number of targets might perform well on synthetic data in which it always satisfies this constraint, but perform poorly if the number of targets varies.

This work presents such an evaluation for a \gls{cnn} trained for target detection and estimation.
We evaluate the detection probability and estimate \gls{rmse} based on measurement data from an outdoor measurement campaign with available groundtruth.
The used \gls{cnn} stems from our previously presented approach for joint detection and estimation of an unknown number of targets in terms of their delay and Doppler-shift, introduced in~\cite{schieler_grid-free_2024}.
\Cref{section:2} presents an overview of the physical narrowband signal model and various random sampling strategies for the synthetic data, as well as the network architecture and training hyperparameters.
\Cref{section:3} introduces the suburban outdoor measurement used for the evaluation with three available links and a single target with available groundtruth.
Finally, the evaluation in \Cref{section:4} includes a qualitative assessment of estimated targets, along with an analysis of detection probability and estimation accuracy based on the available ground truth data.
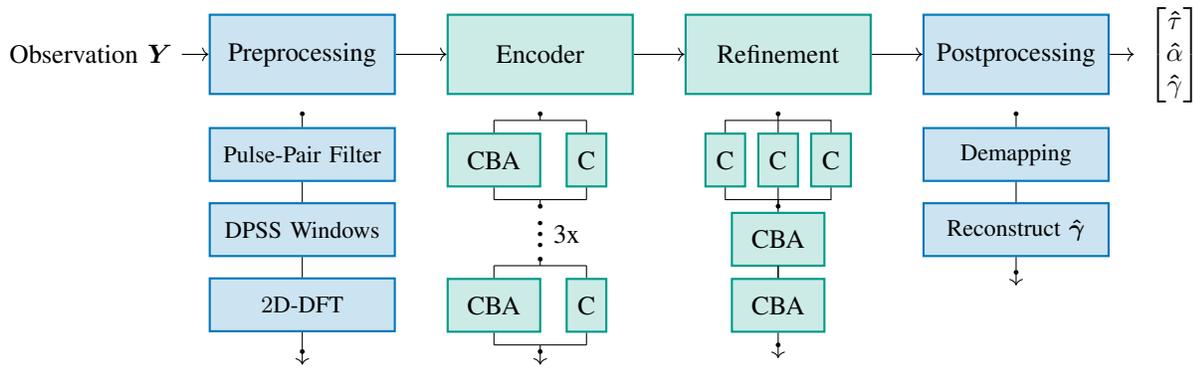
\begin{figure*}[t]
    \centering
    \begin{scaletikzpicturetowidth}{\textwidth}
        % TiKZ style file generated by TikZiT. You may edit this file manually,
% but some things (e.g. comments) may be overwritten. To be readable in
% TikZiT, the only non-comment lines must be of the form:
% \tikzstyle{NAME}=[PROPERTY LIST]

% title typeface
\newcommand{\titleface}[1]{\normalsize{\textbf{#1}}}
\def\blockdist{9em}
\def\boxwidth{7em}
\def\boxheight{3em}

\tikzstyle{dp upper}=[minimum width=\boxwidth, minimum height=\boxheight, align=center, draw=QC1, fill=QC1!20, thick, anchor=center]
\tikzstyle{dp lower}=[minimum width=\boxwidth, minimum height=0.75*\boxheight, align=center, draw=QC1, fill=QC1!20, thick, anchor=center]
\tikzstyle{dp fc}=[minimum width=0.5*\boxwidth, minimum height=0.75*\boxheight, align=center, draw=QC3, fill=QC3!20, thick, anchor=center]
\tikzstyle{dp cba}=[minimum width=0.5*\boxwidth, minimum height=0.75*\boxheight, align=center, draw=QC3, fill=QC3!20, thick, anchor=center]
\tikzstyle{dp sum}=[fill=black, draw=black, shape=circle, inner sep=0pt, minimum size=0.1em, align=center, thick, anchor=center]

\tikzset{
    pics/cba/.style n args={3}{%
        code={%
            \node[inner sep=0pt] (#1) at (0em,1.5em) {};
            \node[minimum width=3.5em, minimum height=2em, align=center, draw=QC3, fill=QC3!20, thick, anchor=center] (CBA) at (0em,0em) {#3};
            \node[inner sep=0pt] (#2) at (0em,-1.5em) {};

            \draw[-] (#1.center) to (CBA.north);
            \draw[-] (CBA.south) to (#2.center);
        }
    },
    pics/cba/.default={In}{Out}{CBA}
}

\tikzset{
  pics/cbaskip/.style n args={3}{% 1 = skip, 2 = skip with C
        code={
            \pic at (-1.75em, 0) {cba={In/CBA}{Out/CBA}{CBA}};
            \node[dp sum, inner sep=0pt] (#1) at (0em, 1.75em) {};
            \node[inner sep=0pt] (P1) at (0em, 1.5em) {};
            \node[inner sep=0pt] (R1) at (1.75em, 1.5em) {};
            \node[inner sep=0pt] (L1) at (-1.75em, 1.5em) {};
            \node[inner sep=0pt] (R2) at (1.75em, -1.5em) {};
            \node[inner sep=0pt] (L2) at (-1.75em,-1.5em) {};
            \node[inner sep=0pt] (P2) at (0em, -1.5em) {};
            \node[dp sum, inner sep=0pt] (#2) at (0em, -1.75em) {};

            % connections
            \draw[-] (#1.center) -- (P1.center);
            \draw[-] (P1.center) -- (R1.center);
            \draw[-] (P1.center) -- (L1.center);
            \draw[-] (P2.center) -- (R2.center);
            \draw[-] (P2.center) -- (L2.center);
            \draw[-] (P2.center) -- (#2.center);
            
            \ifthenelse{#3=1}{
                \draw[-] (R1.center) to (R2.center);
            }{
                \node[minimum width=1.5em, minimum height=2em, align=center, draw=QC3, fill=QC3!20, thick, anchor=center](C) at (1.75em,0em) {C};
                \draw[-] (R1.center) to (C.north);
                \draw[-] (C.south) to (R2.center);
            }
        }
    },
    pics/cbaskip/.default={In}{Out}{1}
}

\tikzset{
  pics/spp/.style n args={2}{
        code={
            \node[dp sum, inner sep=0pt] (#1) at (0em,1.75em) {};
            \node[inner sep=0pt] (P1) at (0em,1.5em) {};
            \node[minimum width=1.5em, minimum height=2em, align=center, draw=QC3, fill=QC3!20, thick, anchor=center] (SPP/1) at (-2em,0) {C};
            \node[minimum width=1.5em, minimum height=2em, align=center, draw=QC3, fill=QC3!20, thick, anchor=center] (SPP/2) at (0,0) {C};
            \node[minimum width=1.5em, minimum height=2em, align=center, draw=QC3, fill=QC3!20, thick, anchor=center] (SPP/3) at (+2em,0) {C};
            \node[inner sep=0pt] (P2) at (0em,-1.5em) {};
            \node[dp sum, inner sep=0pt] (#2) at (0em,-1.75em) {};        
    
            \draw[-] (#1.center) -- (P1.center);
            \foreach \x in {1,...,3}{
                \draw[-] (P1.center) -| (SPP/\x.north);
                \draw[-] (SPP/\x.south) |- (P2.center);
            }
            \draw[-] (P2.center) -- (#2.center);
        }
    },
    pics/spp/.default={In}{Out}
}

\tikzset{
  pics/blockset/.style n args={4}{
        code={
            \foreach \x [count=\xi from 1] in {#3}{
                \node [style=#4, anchor=north, minimum height=2em] (Pre/\xi) at (0,0)[shift={(0em,-(\xi-1)*\boxheight)}] {\small{\x}};
                \xdef\rememberxi{\xi}  % to recall xi outside the loop
            }
            
            % we need xi-1 arrows, so subtract -1 from xi
            \edef\xi{\rememberxi}  
            \pgfmathsetmacro{\xi}{int(\xi-1)}

            \node[dp sum, inner sep=0pt] (#1) at (0em,0.5em) {};
            \node[dp sum, inner sep=0pt] (#2) at (0em,0em)[xshift=0em,yshift=-(\xi+1)*\boxheight+0.5em] {};
            \draw[-] (#1) to (Pre/1.north);
            \pgfmathsetmacro{\xo}{int(\xi+1)}
            \draw[-] (Pre/\xo.south) to (#2);
            % iterate to draw arrows
            \foreach \x in {1,...,\xi}{
                \pgfmathsetmacro{\y}{int(\x+1)}
                \draw[-] (Pre/\x.south) to (Pre/\y.north);
            }
        }
    },
    pics/spp/.default={In}{Out}{{A, B, C, D}}{dp lower}
}

\tikzset{
    pics/vdots/.style n args={1}{
        code={
            \node[] at (-0.0em, 0em) {$\boldsymbol{\vdots}$};
            \node[] at (+1em, -0.3em) {#1};
        }
    },
    pics/vdots/.default={N}
}

\begin{tikzpicture}
	\begin{pgfonlayer}{nodelayer}
        % upper 
        \node [style=dp upper, minimum width=0em, align=center, draw=none, fill=none, thick] (D) at (0*\blockdist,0)[xshift=1em,yshift=0em] {Observation $\bm Y$};
        \node [style=dp upper, draw=QC1, fill=QC1!20, thick] (Pre) at (1*\blockdist,0)[xshift=0em,yshift=0em] {Preprocessing};
        \node [style=dp upper, draw=QC3, fill=QC3!20, thick] (Enc) at (2*\blockdist,0)[xshift=0em,yshift=0em] {Encoder};
        \node [style=dp upper, draw=QC3, fill=QC3!20, thick] (Eta) at (3*\blockdist,0)[xshift=0em,yshift=0em] {Refinement};
        \node [style=dp upper, draw=QC1, fill=QC1!20, thick] (Pos) at (4*\blockdist,0)[xshift=0em,yshift=0em] {Postprocessing};
        \node [style=dp upper, draw=none, fill=none, thick] (Est) at (5*\blockdist,0)[xshift=-3em,yshift=0em] {%
            $\begin{bmatrix}
                \bm\hat{\tau} \\
                \bm\hat{\alpha} \\
                \bm\hat{\gamma}
            \end{bmatrix}$%
        };
%        \node [style=dp upper, draw=QC3, fill=QC3!20, thick] (Mo) at (5.3*\blockdist,0)[xshift=0em,yshift=0em] {Modelorder};

        % lower: 1
        \pic at (1*\blockdist, -2.75em) {blockset={{In1}{Out1}{Pulse-Pair Filter, DPSS Windows, \num{2}D-DFT}{dp lower}}};
        \draw[->] (Out1.center) to ([shift={(0em,-0.5em)}]Out1);

        % lower: 2
        \pic at (2*\blockdist, -3em)[shift={(0em, -1.0em)}] {cbaskip={In21}{Out21}{2}};
        \pic at (2*\blockdist, -5.5em)[shift={(0em, -1.0em)}] {vdots={3x}};
        \pic at (2*\blockdist, -8.5em)[shift={(0em, -1.0em)}] {cbaskip={In23}{Out23}{2}};
        \draw[->] (Out23.center) to ([shift={(0em,-0.5em)}]Out23);

        % lower: 3
        \pic at (3*\blockdist, -4.5em)[shift={(0em, 0.5em)}] {spp={In31}{Out31}};
        \pic at (3*\blockdist, -7.5em)[shift={(0em, 0.5em)}] {cba};
        \pic at (3*\blockdist, -10em)[shift={(0em, 0.5em)}] {cba={In33}{Out33}{CBA}};
        \node[dp sum, inner sep=0pt] at (Out33.center) {};
        \draw[->] (Out33.center) to ([shift={(0em,-0.5em)}]Out33);

        % lower: 4
        \pic at (4*\blockdist, -2.75em) {blockset={{In4}{Out4}{Demapping, Reconstruct $\bm{\hat{\gamma}}$}{dp lower}}};
        \draw[->] (Out4.center) to ([shift={(0em,-0.5em)}]Out4);

	\end{pgfonlayer}
	\begin{pgfonlayer}{edgelayer}
        % upper
        \draw [->] (D.east) to (Pre.west);
        \draw [->] (Pre.east) to (Enc.west);
        \draw [->] (Enc.east) to (Eta.west);
        \draw [->] (Eta.east) to (Pos.west);
        \draw [->] (Pos.east) to ([xshift=2em]Est.west);
	\end{pgfonlayer}
\end{tikzpicture}
    \end{scaletikzpicturetowidth}
    \caption{Network architecture with preprocessing, trainable blocks, and postprocessing. Lower parts detail the steps in the upper part.}
    \label{fig:architecture}
\end{figure*}

\section{Our Approach}\label{section:2}
We employ the same approach as \cite{schieler_grid-free_2024}, that is, we use a \gls{cnn} to detect and estimate the parameters of an unknown number of targets.
This section provides a brief summary of the signal model, describes the changes to the preprocessing compared to \cite{schieler_grid-free_2024}, and lists training hyperparameters.
\subsection{Signal Model}
For our work, we consider a typical outdoor \gls{isac} propagation scenario.
We assume that the wireless channel between \gls{tx} and \gls{rx}, in which the transmitted signal is reflected, scattered, and diffracted, can be modeled by multiple propagation paths representing moving targets and static clutter.
Each path represents a planar wave that impinges on the \gls{rx}, which is characterized by its propagation delay $\tau_p \in \R$, its Doppler-shift $\alpha_p \in \R$, and its complex magnitude $\gamma_p \in \C$.
The voltage measured by \gls{rx} at each antenna element is directly proportional to the electric field strength created by the superposition of the planar wavefronts.
Hence, the overall wireless channel can be modeled as a \gls{ltv} system composed of $1..P \in \N$ specular paths.
The static clutter paths have $\alpha_p = \qty{0}{\hertz}$ while paths corresponding to targets have $\alpha_p \neq \qty{0}{\hertz}$.
Our task is to detect and estimate the target parameters of such a \gls{siso} channel.

We assume the \gls{rx} continuously samples the received signal, whose sampled time-variant channel transfer function of bandwidth $B$ can be modeled under the narrowband assumption ($B \ll f_c$). 
The sampling process involves sampling intervals in frequency $\Delta f > 0$ and time $\Delta t > 0$, where the signal $\bm S$ is sampled $N_f$ times in frequency and $N_t$ times in time. The sampling points are given by:
\begin{equation}
    f_k = f_0 + k \cdot \Delta f \quad \text{and} \quad t_l = t_0 + l \cdot \Delta t,
\end{equation}

where $k = 0, \ldots, N_f - 1$, $l = 0, \ldots, N_t - 1$, $f_0 = -\frac{B}{2}$ and $t_0 = 0$.
The noiseless, sampled observation of the time-variant channel transfer function, composed of $P$ paths, is denoted as
\begin{align}\label{eq:fwdmodel}
    H_{k,l}(\bm \gamma, \bm \tau, \bm \alpha) = \sum_{p=1}^{P} \gamma_p \exp{(-2j\pi f_k \tau_p)} \exp{(2j\pi t_l \alpha_p)},
\end{align}
where $\theta_p = \{\gamma_p, \tau_p, \alpha_p\}$ denotes the parameters of the $p$-th path. 

The noisy observation is denoted by 
\begin{equation}\label{eq:observation}
    \bm Y = \bm H + \bm N,
\end{equation}
where $\bm N \in \C^{N_f, N_t}$ represents a complex, zero-mean, uncorrelated Gaussian noise process with element-wise variance $\sigma^2$.
In an \gls{isac} system, the \gls{rx} obtains \eqref{eq:observation} from channel estimation techniques, such as zero-forcing \cite{thoma_cooperative_2019}.
For simplicity, we use $\eta_p$ to describe only the non-linear parameters such that $\theta_p = \{\gamma_p \quad \eta_p\}$.
The objective of our \gls{cnn} is to identify the sensing targets in $\bm Y$ and provide an estimate of $\hat{\eta_p}$ for each path.
To recover the linear signal parameters $\hat{\gamma}$, we use the least-squares method.
\begin{table}[b]
    \centering
    \small
    \caption{Dataset summary and training hyperparameters.}
    \label{tab:settings}
    \begin{tabularx}{\linewidth}{Xp{4.5cm}}
        \toprule
        Name                                 & Value                                                                                  \\ \midrule
        \textbf{Datasets} & \mbox{}                                                                                                   \\
        Distribution $\tau_p$, $\alpha_p$    & $\mathfrak{U}_{[0,\tau_{\text{max}}]}$ and $\mathfrak{U}_{[-\alpha_{\text{max}},\alpha_{\text{max}}]}$                            \\
        Number of Samples                    & $N_f=\num{1024}$, $N_t=\num{100}$                                                      \\
        Input Data $\bm Y_1$                 & $6 \times 512 \times 512$ ($2N_w \times N_{\tau} \times N_{\alpha}$)                                   \\
                                             & $\tau_{\text{max}} = 0.02$, $\alpha_{\text{max}} = 0.05$ \\ 
        Magnitudes                           & $\mathfrak{U}_{[0.001, 1]}$                                                            \\
        Phases                               & $\mathfrak{U}_{[0,2\pi]}$                                                              \\
        SNR                                  & \qtyrange{0}{50}{\decibel}                                                             \\
        Number of Paths                      & $\mathfrak{U}_{[1,10]}$                                                                \\
        Trainingset Size                     & \num[exponent-mode=engineering, drop-zero-decimal=true]{200000}                        \\ \midrule
        \textbf{Training}                    &                                                                                        \\
        Optimizer                            & Adam~\cite{kingma_adam_2014}, $\gamma=0.0003$, \newline $\beta_1=0.9$, $\beta_2=0.999$ \\
        Mini-Batchsize                       & \num{512}                                                                              \\
        Epochs                               & 100                                                                                    \\
        Trainable Parameters                 & \num[exponent-mode=engineering, drop-zero-decimal=true]{1.3e6} \\\bottomrule
    \end{tabularx}
\end{table}
\subsection{Preprocessing and Architecture}
We use the architecture presented in \cite{schieler_grid-free_2024}, which is illustrated in \Cref{fig:architecture}.
It consists of four main blocks: a deterministic (read untrained) preprocessing, two trainable blocks called encoder and downsampling, and a deterministic postprocessing.

The task of deterministic preprocessing is to reshape the input data into useful features for \gls{cnn}. 
Being deterministic, it allows chaining different signal processing steps, i.e., multi-windowing, correlation, resampling, and normalization, to enable scenario-specific adaptations.
For example, as we expect a multipath-rich propagation environment with strong \gls{los} and static clutter, we employ a \gls{ppf}.
Included the \gls{ppf} during training, the \gls{cnn} learns to compensate for its filter transfer function on the data.

As \glspl{cnn} can process multiple channels in parallel (similar to an RGB image), we also take advantage of strategies from multitaper analysis~\cite{thomson_spectrum_1982}.  
To this end, we employ orthogonal \gls{dpss} windows to reduce bias and spectral leakage.
We parametrize them to a standardized half-bandwidth of $NW=2$, resulting in $N_w = 3$ different \gls{dpss} windows.
Finally, the multi-window representations are first zero-padded, then converted into the delay-Doppler domain using a two-dimensional discrete Fourier transform (2D-DFT), and finally cropped to retain only the relevant delay and Doppler-shift ranges (see \Cref{tab:settings}). 

Finally, the final preprocessing step maps the complex samples to the real number space. 
For this, we use $f_1:x \mapsto \log_{10}(x)$ and $f_2: x \mapsto \angle x$. 
We stack the results in the channel dimension, such that we obtain $2N_w = 6$ channels for the \gls{cnn} input data $\bm Y_1$.
The remaining steps for the \gls{cnn} and postprocessing follow the ones presented in \cite{schieler_grid-free_2024}.
\begin{figure}[t]
    \def\mathdefault#1{#1}
    \begin{center}
        \input{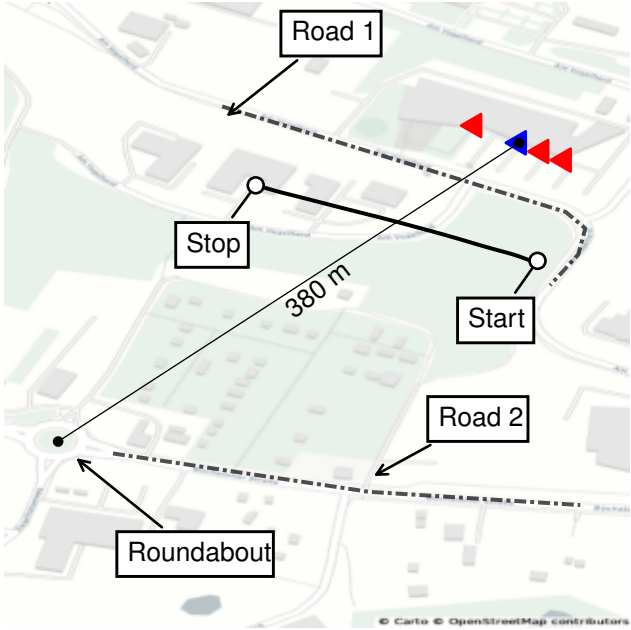}
    \end{center}
    \caption{Measurement scenario. A basestation uses a \gls{tx} (red) and three distributed \gls{rx} (blue) for sensing the coverage area in front of an industrial building. In addition to a known \gls{uav} flight path, automotive targets-of-opportunity are present on nearby roads.}
    \label{fig:scenario}
\end{figure}
\subsection{Synthetic Dataset}
For applicability to the measurement data during the inference phase, the synthetic data must be sufficiently similar.
Therefore, we maintain the following assumptions about the apriori knowledge for each observed snapshot $\bm Y$
\begin{enumerate}
    \item the number of paths $P$ is independent and unknown 
    \item the set of paths $\theta$ is independent and unknown
    \item the paths may be insufficiently separated, i.e., no minimum separation is enforced
    \item the noise level $\sigma^2$ is independent and unknown
\end{enumerate}

To support them, we employ different random sampling processes to generate the parameters required to compute a synthetic observation $\bm Y$.
First, we sample $P$ from a uniform distribution from \num{1} to \num{30} to reflect the unknown number of paths and account for the multipath rich scenario. 
Secondly, each component of $\theta$ is sampled from a different random distribution for each path, as specified in \Cref{tab:settings}.
We chose a uniform distribution for the delay and Doppler-shifts, with the limits specified in \Cref{tab:settings}.
The delay limits reflect the \gls{ofdm}-specific limitations for \gls{ofdm}, where the limit for the maximum excess delay must not exceed the \gls{cp} length.
The limits for the Doppler-shift are based on the comparably small velocities expected in the propagation scenario. 
These limits emphasize that discernible propagation paths in the scenario are limited to the beginning of the channel impulse response and small bistatic velocities.
The real and imaginary parts of the complex path weights $\gamma$ are derived from separate random sampling processes for magnitude and phase.
Finally, with the full set of $\theta$, we compute \cref{eq:fwdmodel} and add a random realization of $\bm N$.
Its noise variance is scaled relative to the total signal power, with respect to a random \gls{snr} from \qtyrange{-30}{50}{\decibel} for each realization.

In summary, each synthetic snapshot generated is composed of a random model order, a different set of parameters, and noise realization.
Perhaps unsurprisingly, our experiments show that this avoids overfitting, as the different sources of randomness make it unlikely to generate the exact same $\bm Y$ twice.
\begin{figure*}
    \def\mathdefault#1{#1}
    \input{images/full/full.tex}
\end{figure*}

\section{Measurement Dataset}\label{section:3}
For evaluation, we used a scenario from the \textit{isac-uav} dataset \cite{beuster_measurement_2023,schieler_dataset_2024}.
It contains measurements of \cref{eq:observation} with multiple \glspl{rx} recording a \gls{uav} flight in a suburban outdoor environment.
The measurement area in the city of Ilmenau, Germany, and the selected scenario are illustrated in \Cref{fig:scenario}.
The \gls{ofdm}-like measurement signal consisted of a Newman-sequence, transmitted at $f_c = \qty{3.75}{\giga\hertz}$, composed of \num{1600} subcarriers, of which \qty{20}{\percent} were used as a guardband, such that $N_f = 1280$.
Along with the symbol duration of \qty{16}{\micro\second}, the total bandwidth is $B = \qty{80}{\mega\hertz}$ and a corresponding delay resolution of \qty{12.5}{\nano\second}.
For integration in the slow-time domain, we selected $N_t = 100$, such that the coherent processing interval equals \qty{32}{\milli\second}, with a Doppler bandwidth of $B_\alpha = \qty{3.125}{\kilo\hertz}$ (with $\Delta t = \qty{320}{\micro\second}$).
At $f_c=\qty{3.75}{\giga\hertz}$, this results in a maximum target velocity of $v_{\text{max}} = \qty{125}{\meter\per\second}$, and hence is well above what is expected in a suburban scenario.

A groundtruth for the \gls{uav} dynamic state vector is provided via an independent \gls{rtk}-based positioning system on the \gls{uav}.
Using it along with the known \gls{tx} and \glspl{rx} positions, we compute the groundtruth for the delay and Doppler-shift parameters of the \gls{uav} analytically for each link.
I.e., the delay is computed as
\begin{equation}
    \tau_p = \frac{1}{c_0} \left( R_{\text{Tx}} + R_{\text{Rx}} \right),
\end{equation}
where $c_0$ is the speed-of-light in air and $R_{Tx}$ and $R_{Rx}$ express the distances between the \gls{tx} and the target, and the target and \gls{rx}, respectively.
The analytical expression for the bi-static Doppler-shift is 
\begin{equation}
  \alpha_p = \frac{2 v f_c}{c} \cos \psi \cos \left( \frac{\beta}{2} \right)
\end{equation}
where $f_c$ denotes the carrier frequency, $v$ the magnitude of the target's velocity vector, the bi-static angle $\beta$, and the angle $\psi$ between the target's velocity vector and the bisector of the bi-static angle.
Due to the suburban environment, other radar sensing targets, such as cars and pedestrians (see \Cref{fig:scenario}), are also part of the measurement data. 
As groundtruths for these targets are unavailable, we refer to them as \textit{targets-of-opportunity} and only qualitatively evaluate their corresponding detections.
More details of the measurement equipment and setup are available in \cite{beuster_measurement_2023}.

\section{Evaluation}\label{section:4}
\begin{figure*}[t]
    \def\mathdefault#1{#1}
    \input{images/zoomed/zoomed.tex}
\end{figure*}
The goal of our work is to see if our approach can successfully detect and estimate target parameters in the measurement data after training on synthetic data.
Such a property is highly desirable, as it alleviates the need for costly, possibly site- or hardware-specific, measurement data collection and manual labeling efforts.
Comparably, synthetic data is cheap to acquire, and is by its very nature, already labeled with a complete groundtruth.

As the chosen dataset contains a dedicated groundtruth only for the target \gls{uav}, we decided on a two-fold evaluation.
First, we check whether our approach transferred successfully from synthetic to measurement by exploiting knowledge about the position (not speed) of the aforementioned \textit{targets-of-opportunity}, as these were found to correspond to vehicles on roads.
Second, a quantitative evaluation, for which we filter the estimates and compute the detection probability and accuracy, using the \gls{rtk} groundtruth of \gls{uav}. 

\subsection{Automotive Targets}\label{sec:eval:qual}
\Cref{fig:full} illustrate the results obtained in the processed measurement scenario for the three \glspl{rx}.
Given $N_s$ non-overlapping snapshots, the background shows the maximum power for each bin as
\begin{equation}
   \vert \bm Y_{\text{max}} \vert = \max_{n \in [1, N_s]} \vert \bm Y[n] \vert.
\end{equation}
Three distinct clusters of targets are well-observable, loosely termed \textit{Road 1}, \textit{Road 2}, and the \gls{uav} track.
The \textit{Road 1} tracks correspond to automotive targets in front of the building with the measurement equipment (see also \Cref{fig:scenario}), with their Doppler-shifts changing significantly while passing the building.
The trajectories of targets on the distant road connected to the roundabout, labeled \textit{Road 2}, cause comparably smaller Doppler-shifts due to their more perpendicular motion to the measurement setup.
Weakly visible is also the track caused by the \gls{uav}, whose groundtruth is plotted as a line with the characteristic start and end points marked in the plots. 

We also plot the point estimates from our approach as blue crosses, but for a clearer figure, limit it to every $50$-th snapshot.  
However, one can clearly observe the estimates visibly aligned with maxima in $\vert Y_{\text{max}} \vert^2$ (or \enquote{target peaks}), indicating a successful estimation in the measurement data.
However, the \gls{uav} is only weakly visible in these plots and it remains unclear if it is detected.
\subsection{Quantitative Results}\label{sec:eval:quan}
\begin{figure*}[t]
    \def\mathdefault#1{#1}
    \input{images/time/time.tex}
\end{figure*}
Although the previous results show our approach works, they lack any relation to the actual measurement time, detection probability, or estimation accuracy. 
To assess those, we zoom in closer to the \gls{uav} flight paths, and filter the estimates with the groundtruth estimates computed from the \gls{uav} positions recorded with the \gls{rtk}.
Let $\eta^{[i]} = [\tau^{[i]} \alpha^{[i]}]^T$ denote the groundtruth delay and Doppler-shift of the \gls{uav} in the $i$-th snapshot.
This \textit{groundtruth filter} is expressed as
\begin{align}
\hat{\eta}_{k^*} = 
\begin{cases}
   \begin{array}{ll}
      \operatorname*{arg\,min}_{\hat{\eta}_k \in \{\hat{\eta}\}} \|\hat{\eta}_k - \eta\|, \text{if } &|\hat{\tau}_{k} - \tau | < \epsilon_\tau \text{ and} \\
      & |\hat{\eta}_{k,2} - \eta_2| < \epsilon_\alpha \\
      \emptyset, & \text{otherwise}
   \end{array}
\end{cases}
\end{align}, 
which returns a single or no estimate (the empty set $\emptyset$), if the closest estimates error exceeds the threshold for either $\epsilon_\tau$ or $\epsilon_\alpha$. 
We chose maximum errors based on the sampling process, such that $\epsilon_{\tau} = \frac{3}{N_f\Delta_f} = \qty{37.5}{\nano\second}$ and $\epsilon_\alpha = \frac{3}{N_t\Delta_t} = \qty{93.75}{\hertz}$.
For an estimate of the detection probability $P_D$, we then compare the number of snapshots without estimates $N_\emptyset$ over the total number of snapshots $N_{\text{meas}}$, as
\begin{equation}\label{eq:detprob}
    P_D = 1 - \frac{N_\emptyset}{N_\text{meas}}.
\end{equation}
We compute the \gls{rmse}, separately for $\tau$ and $\alpha$, only for the detections sufficiently close to the target groundtruth.

\Cref{fig:zoom} shows only the groundtruth filtered estimates over the full measurement time of \qty{32}{\second}, while \Cref{fig:time} plots the same estimates separately in delay and Doppler-shift over the measurement time. 
The results confirm the previous findings that our approach is well applicable to the measurement data, since the estimates $\hat{\tau}$ and $\hat{\alpha}$ visibly align with the corresponding groundtruth values.
The strong drop in detection probability (and target return) is clearly visible at the beginning and end of the measurement. 
This is due to the directivity of the antennas used, as the \gls{uav} enters and leaves the main beams.
Furthermore, the detection probability and \gls{rmse} in each \gls{rx} are shown in \Cref{tab:results}.
The results \gls{rmse} highlight that the approach is partially capable of super-resolution and can detect and estimate the sensing target.
\begin{table}[t]
    \centering
    \small
    \caption{Evaluation results for the detection probability $P_{D}$ the \gls{rmse} in delay and Doppler-shift.}
    \label{tab:results}
    \begin{tabularx}{\linewidth}{XXXX}
        \toprule
             & $P_{D}$  & \gls{rmse} $\sigma_{\tau}$ & \gls{rmse} $\sigma_{\alpha}$  \\ \midrule
        Rx 1 & 0.54     & \qty{16.2}{\nano\second} & \qty{10.4}{\hertz} \\
        Rx 2 & 0.48     & \qty{17.2}{\nano\second} & \qty{9.8}{\hertz} \\
        Rx 3 & 0.55     & \qty{18.2}{\nano\second} & \qty{7.9}{\hertz} \\ \bottomrule
    \end{tabularx}
\end{table}

\section{Conclusion}\label{section:5}
This paper presents the successful application of a synthetically trained \gls{cnn} to measurement data, demonstrating effective detection and estimation of a small \gls{uav} and other targets.
This highlights that our approach is well-suited for radar sensing systems, including future \gls{isac} systems.
Future studies should include detailed performance comparisons with other detection and estimation methods, such as \gls{cfar} for detection and iterative maximum likelihood for estimation.
Additionally, the aforementioned impacts of signal sparsity must be carefully analyzed.
A promising extension of this approach is the integration of target classification capabilities, where \glspl{cnn} are already applied (e.g., see \cite{youssef_scalable_2023}).
In particular, spectral features caused by extended targets or Micro-Doppler effects could be leveraged to achieve joint detection, estimation, and classification, further advancing the adaptability and robustness of \gls{ml}-based radar sensing.

\section*{Acknowledgment}\label{section:6}
The authors acknowledge the financial support by the Federal Ministry of Education and Research of Germany in the project “Open6GHub” (grant number: 16KISK015), and “KOMSENS-6G” (grant number: 16KISK125).
\vfill

\printbibliography

\end{document}